\documentclass[twocolumn,showpacs,preprintnumbers,amsmath,amssymb]{revtex4}
\usepackage{graphicx}% Include figure files
\usepackage{dcolumn}% Align table columns on decimal point
\usepackage{bm}% bold math

\begin{document}

\preprint{}

\title{Spin-Related Current Suppression in a Semiconductor-Quantum-Dot  Spin-Diode Structure}% Force line breaks with \\

\author{K. Hamaya,$^{1,2,3,4}$\footnote{E-mail: hamaya@ed.kyushu-u.ac.jp} M. Kitabatake,$^{1}$ K. Shibata,$^{1}$ M. Jung,$^{1}$ S. Ishida,$^{5}$ T. Taniyama,$^{4,6}$\\ K. Hirakawa,$^{1,2,7}$ Y. Arakawa,$^{1,2}$ and T. Machida$^{1,2,7}$\footnote{E-mail: tmachida@iis.u-tokyo.ac.jp}}
\affiliation{% 
$^{1}$Institute of Industrial Science, University of Tokyo, 4-6-1 Komaba, Meguro-ku, Tokyo 153-8505, Japan}%
\affiliation{% 
$^{2}$Institute for Nano Quantum Information Electronics, University of Tokyo, 4-6-1 Komaba, Meguro-ku, Tokyo 153-8505, Japan}%
\affiliation{% 
$^{3}$Department of Electronics, Kyushu University, 744 Nishi-ku, Fukuoka 819-0395, Japan.}%
\affiliation{%
$^{4}$Japan Science and Technology Agency, PRESTO, 4-1-8 Honcho, Kawaguchi 332-0012, Japan}%
\affiliation{%
$^{5}$Research Center for Advanced Science and Technology, University of Tokyo, 4-6-1 Komaba, Meguro-ku, Tokyo 153-8505, Japan}%
\affiliation{%
$^{6}$Materials and Structures Laboratory, Tokyo Institute of Technology, 4259 Nagatsuta, Midori-ku, Yokohama 226-8503, Japan}
\affiliation{%
$^{7}$Japan Science and Technology Agency, CREST, 4-1-8 Honcho, Kawaguchi 332-0012, Japan}%
%\affiliation{
%%Second institution and/or address\\

\date{\today}% It is always \today, today,
\begin{abstract}
We experimentally study the transport features of electrons in a spin-diode structure consisting of a single semiconductor quantum dot (QD) weakly coupled to one nonmagnetic (NM) and one ferromagnetic (FM) lead, in which the QD has an artificial atomic nature. A Coulomb stability diamond shows asymmetric features with respect to the polarity of the bias voltage. For the regime of two-electron tunneling, we find anomalous suppression of the current for both forward and reverse bias. We discuss possible mechanisms of the anomalous current suppression in terms of spin blockade via the QD/FM interface at the ground state of a two-electron QD. 
\end{abstract}
%\pacs{73.43.-f}% PACS, the Physics and Astronomy
                    %display DEGired
\maketitle
%\section{INTRODUCTION}
The control of spin degree of freedom in semiconductor-quantum-dot (QD) systems, i.e., artificial atoms and molecules, has been widely studied for future spin-based quantum information processing techniques and semiconductor-based spintronic applications.\cite{Tarucha,Kouwenhoven,Gordon,Petta,Koppens,Fujisawa} Employing a weakly coupled double QD system, Ono {\it et al}. achieved that one QD works as a spin injector with fully spin-polarized electrons and another acts as a spin filter, in which the electron transport from the $|$1$s$, $\uparrow >$ state in the injector to the $|$1$s$, $\downarrow >$ state in the filter is suppressed significantly, i.e., spin blockade based on Pauli exclusion.\cite{Ono} On the other hand, even for single QD systems, QD spin filters have also been proposed,\cite{Recher} and Hanson {\it et al}. have demonstrated a bipolar spin filter using a few-electron QD and its Zeeman splitting.\cite{Hanson} By combining the nature of QDs with spin-polarized electrons injected from ferromagnetic (FM) leads, further interesting phenomena arise.\cite{Recher,Rud,Weymann,Souza} When spin-polarized current flows through a QD state, the tunneling of electrons can be affected by spin conservation. When the spin orientation of electrons in a QD state is opposite to that of the FM lead, the spin transport between them is restricted. As a result, current suppression can occur. For asymmetric nonmagnet (NM)--QD--FM systems, i.e., a spin-diode structure with a semiconductor QD, strong current suppression has been predicted theoretically,\cite{Recher,Rud,Weymann,Souza} giving rise to diode-like current--voltage characteristics. However, due to the difficulties in fabricating semiconductor-QD spin-diode structures, no precise experimental evidence has been found for current suppression associated with the spin states on the QD and the spin-polarized electrons. 

Recently, we combined a {\it single} self-assembled semiconductor (InAs) QD with FM metal (Ni or Co) source--drain reservoirs by fabricating lateral spin-valve devices.\cite{Hamaya,Hamaya2,Hamaya3} Clear Coulomb blockade characteristics were demonstrated even for FM leads at low temperatures. In addition, a tunneling magnetoresistance (TMR) effect originating from spin transport through the single semiconductor QD was observed and could be tuned by the gate voltage.\cite{Hamaya,Hamaya2} We also found that the TMR effect depends on the Coulomb blockade characteristics and discrete energy levels in a QD.\cite{Hamaya3} Furthermore, such devices can realize experimental detection of the Kondo effect between electrons in the QD and FM leads.\cite{Hamaya4} Overall, these results are largely consistent with theoretical predictions.\cite{Weymann,Brataas,Imamura,Barnas2,Jan,Jan2} Hence, this laterally fabricated QD device structure is an ideal system to experimentally study the theoretical predictions. To explore the feasibility of a rectifier for spin current and other devices, we should fabricate a spin-diode structure using our device concepts\cite{Hamaya,Hamaya2,Hamaya3,Hamaya4} and investigated their transport features.
\begin{figure}[t]
\includegraphics[width=8cm]{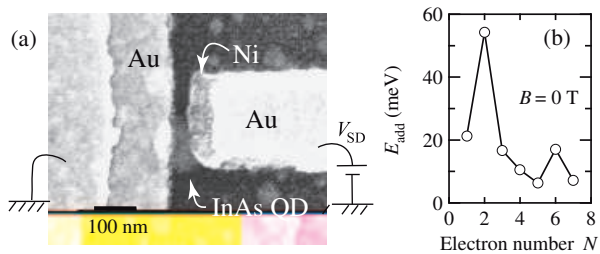}
\caption{(a) Scanning electron micrograph of the lateral NM--QD--FM system. The dot size is $\sim$ 130 nm. (b) Addition energy of electrons vs. electron number for the QD used in this study.}
\end{figure} 

In this Letter, we present the first measurements of electron transport properties for a semiconductor-QD spin-diode structure. A few-electron QD with shell-dependent addition energy can be obtained, and we can experimentally examine correlations between the electron transport features and orbital states in the QD spin-diode structure. Asymmetric current--voltage characteristics are observed with respect to the polarity of the bias voltage. For the regimes of two-electron tunneling, we find anomalous suppression of the current for both forward and reverse bias. We discuss possible origins of the anomalous current suppression in terms of spin blockade for electron transport in the NM--QD--FM system.

Self-assembled InAs QDs were grown on a substrate made of 170-nm-thick GaAs buffer layer/90-nm-thick AlGaAs insulating layer/$n$$^{+}$-GaAs(001). The $n$$^{+}$-GaAs(001) is utilized as an electrode for backgate voltage ($V$$_\mathrm{G}$). On this wafer, we fabricated many nanogaps consisting of wire structures for source--drain electrodes of the QD by conventional electron-beam lithography. Using shadow evaporation techniques and a lift-off method, we produced NM metal (Au/Ti)--QD--FM metal (Ni) nano-junctions as shown in Fig. 1(a). Detailed schematic diagrams of a similar device structure for FM--QD--FM systems have been described in our previous works.\cite{Hamaya,Hamaya3} To get a single domain structure for the FM lead, we used a long wire shape of 200 nm $\times$ 20 $\mu$m. All the transport measurements were performed by a dc method in a $^{3}$He--$^{4}$He dilution refrigerator at 50 mK. External magnetic fields ($B$) were applied parallel to the long axis of the wires in the film plane.
\begin{figure}[t]
\includegraphics[width=7cm]{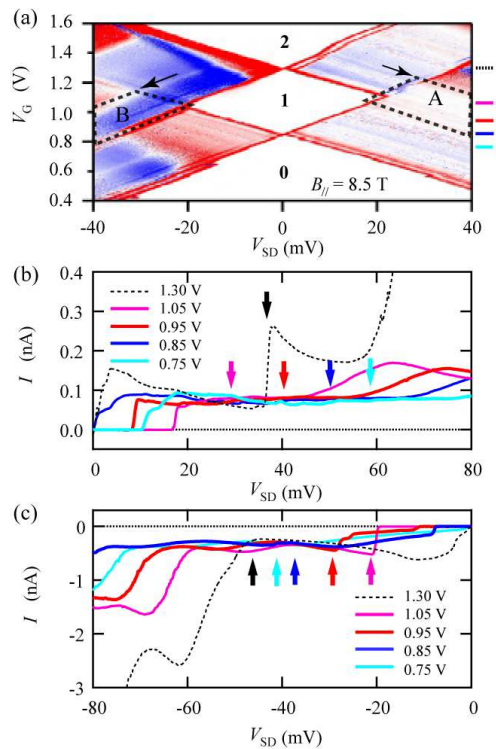}
\caption{(Color online) (a) Differential conductance $dI$/$d$$V$$_\mathrm{SD}$ as a function of $V$$_\mathrm{SD}$ and $V$$_\mathrm{G}$ measured at 50 mK. The red and blue regions are the $dI$/$d$$V$$_\mathrm{SD} >$ 0 and $dI$/$d$$V$$_\mathrm{SD} <$ 0 regimes, respectively. The forward and reverse bias show the electron transport features from the FM to the QD and from the QD to the FM, respectively. (b) and (c) are representative $I$--$V$$_\mathrm{SD}$ characteristics at $V$$_\mathrm{G} =$ 0.75, 0.85, 0.95, 1.05, and 1.30 V in the forward and reverse bias regimes, respectively.}
\end{figure}

We first measured the current ($I$) as a function of $V$$_\mathrm{G}$ at a source--drain bias voltage ($V$$_\mathrm{SD}$) of 100 $\mu$V at $B =$ 0 T. With increasing $V$$_\mathrm{G}$ from 0 V, Coulomb oscillation peaks up to 7--8 were observed before the marked leak current was detected; each electron is added to the QD at the $V$$_\mathrm{G}$ of the Coulomb peak one by one, i.e., it functions as a single electron transistor. There is no conduction ($I \leqslant$ 0.1 pA) with decreasing $V$$_\mathrm{G}$ from 0 V. Hence, we assigned an absolute electron number ($N$) in the QD. Figure 1(b) shows the addition energy ($E_\mathrm{add}$) vs. $N$. We can see clear peaks at $N =$ 2 and 6, and can attribute the two peaks to the shell-dependent energy change of the QD, as previously shown in a vertical QD system with two-dimensional harmonic potential walls.\cite{Tarucha} The level degeneracies of the $s$ and $p$ shells in the QD are identified as the peaks at $N =$ 2 and 6, respectively. The charging energy of the $s$ state is $\sim$ 21 meV and the energy difference between the $s$ and $p$ orbitals is estimated to be $\sim$ 33 meV, consistent with previous work for Au/Ti$-$InAs QD$-$Ti/Au systems.\cite{Jung}

Figure 2(a) shows the differential conductance ($dI$/$d$$V$$_\mathrm{SD}$) as a function of $V$$_\mathrm{SD}$ and $V$$_\mathrm{G}$, i.e., a Coulomb stability diamond (CD), in the few-electron QD regime, measured at $B =$ 8.5 T. This is the first experimental transport measurement for semiconductor-QD spin-diode structures. The white regions labeled by the numbers 0, 1, and 2, indicate the Coulomb blockade regions. Large asymmetries of the conduction changes can be seen with respect to the direction of $V$$_\mathrm{SD}$, and regions with $dI$/$d$$V$$_\mathrm{SD} <$ 0 (blue regions), i.e., negative differential conductance (NDC), are exhibited markedly in the reverse bias regime. Here, we measured CD for various $B$, and clearly observed Zeeman energy splitting $\Delta E$$_\mathrm{Z}$ $=$ g$\mu_{B}B$ ($\mu_{B}$ is Bohr's magneton) and the relevant excited states for both bias regimes.\cite{ref} Similar features for a few-electron QDs have been reported by Hanson {\it et al}.\cite{Hanson} Since the observed $\Delta E$$_\mathrm{Z}$ depended linearly on $B$ (see-Appendix, Fig.5), we roughly determined an effective g-factor of the QD, $|$g$|$$=$ 4.65.\cite{Igarashi}
\begin{figure}[t]
\includegraphics[width=7cm]{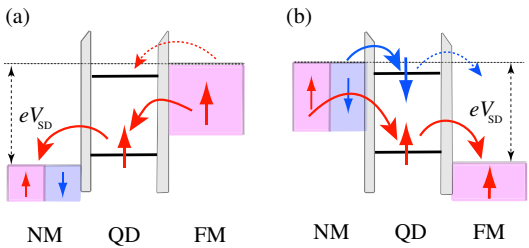}
\caption{(Color online) Schematic diagrams of spin blockade under (a) forward and (b) reverse bias. For simplicity, the FM electrodes are illustrated as HMFs. }
\end{figure} 

Firstly, we note that although region A, enclosed by a black dashed line in the forward bias regime, includes transitions from the one-electron tunneling to the two-electron tunneling, we cannot see any evident conductance changes showing diamond borderlines. To further examine this anomaly, we display representative $I$--$V$$_\mathrm{SD}$ characteristics for various $V$$_\mathrm{G}$ in Fig. 2(b) toward higher $V$$_\mathrm{SD}$. The black dashed curve shows the data measured at $V$$_\mathrm{G} =$ 1.30 V. This data includes the linear transport regime, exhibiting a $N =$ 1$\leftrightarrow$2 transition at $V$$_\mathrm{SD} \sim$ 0 V as well as two Coulomb staircases. The current jump at $V$$_\mathrm{SD} \sim$ 38 mV represents the onset of the regime of two-electron tunneling where the current originates from the fluctuation of $N$ in the QD, with $N =$ 0, 1, or 2. The $I$--$V$$_\mathrm{SD}$ curves for 0.75 V $\lesssim$ $V$$_\mathrm{G}$ $\lesssim$ 1.05 V show different features in the relevant $V$$_\mathrm{SD}$ regime. If a diamond borderline was seen in Fig. 2(a), we should observe current jumps at the arrows depicted in Fig. 2(b), where the color of the arrows corresponds to those of the data. In short, the current jumps showing the two-electron tunneling seem to be suppressed in the region A. We confirmed the suppression of the current jumps at the diamond borderline in the region A even at zero-field (see-Appendix, Fig.6), where the magnetic moments of the wire-shaped FM lead were aligned to the long axis of the wire because of its strong shape anisotropy.
\begin{figure}[t]
\includegraphics[width=6.5cm]{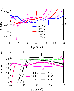}
\caption{(Color online) Enlarged $I$--$V$$_\mathrm{SD}$ characteristics for various $V$$_\mathrm{G}$ in (a) forward and (b) reverse bias. }
\end{figure}   

Next, we consider the region B in the reverse bias regime. The $I$--$V$$_\mathrm{SD}$ characteristics including the relevant region are shown in Fig. 2(c). The black dashed curve is the same data as shown in Fig. 2(b) ($V$$_\mathrm{G} =$ 1.30 V). With increasing negative $V$$_\mathrm{SD}$, we can see Coulomb staircase-like current changes. The current increases significantly at $V$$_\mathrm{SD} \sim$ $-$ 45 mV, which also shows the onset of two-electron tunneling. For the data for 0.75 V $\lesssim$ $V$$_\mathrm{G}$ $\lesssim$ 1.05 V, however, the current increase showing the onset of two-electron tunneling cannot be found at the arrows depicted in Fig. 2(c). From these results, we conclude that there are mechanisms restricting two-electron tunneling in the two-electron QD regime for our QD spin-diode structure. Such restricted current flow could not be seen in previous Au/Ti$-$InAs QD$-$Ti/Au systems.\cite{Jung}

To discuss the anomalous suppression of the current jumps due to the onset of two-electron tunneling, we take into account spins on the QD. Figures 3(a) and (b) show schematic diagrams of spin transport expected for the two-electron tunneling regime under a forward bias and a reverse bias, respectively. For simplicity, we illustrate the case using a half-metallic ferromagnet (HMF) as the FM lead. We assume that the spin relaxation time is longer than the typical time it takes an electron to tunnel into and out of the QD. As the electrons are injected from the FM lead to the QD at a forward bias, the tunneling rate of the electrons depends on the spin states in the QD because the electrons in the FM lead are spin polarized. For the two-electron QD regime, in general, the 1$s$ orbital level has a ground state of ($\uparrow$, $\downarrow$), a spin singlet and total spin $S =$ 0,\cite{Tarucha,Kouwenhoven} as confirmed by the artificial atomic nature shown in Fig. 1(b). Thus, the tunneling rate of the spin-up electrons from the FM lead into the $|$1$s$, $\downarrow >$ state is generally slow, based on spin selective tunneling. Namely, the spin blockade in the QD state can be inferred [Fig. 3(a)], causing current suppression. For reverse bias, a different scenario can be considered. Since the transport of electrons from the NM lead to the QD is not limited, both spin-up and spin-down electrons can enter the QD states, $|$1$s$, $\uparrow >$ and $|$1$s$, $\downarrow >$, leading to a spin singlet for the two-electron QD regime. For spin transport from the QD to the FM lead, however, the tunneling rate of the spin-down electrons is slower than that of the spin-up electrons because the band structure of the FM lead at the Fermi level is spin polarized. Once the $|$1$s$, $\downarrow >$ in the QD is occupied with the spin-down electrons, it can be difficult to tunnel out of the QD through the barrier [Fig. 3(b)], i.e., single spin saturation occurs.\cite{Rud,Weymann,Souza} As a consequence, the next spin-down electrons cannot enter the $|$1$s$, $\downarrow >$ state in the QD, so that the current is also suppressed. This phenomenon is also one contribution of the spin blockade.  

We concentrate again on the forward bias data to further examine the anomalous suppression of the current in detail. Figure 4 (a) displays an enlarged figure of the $I$--$V$$_\mathrm{SD}$ characteristics for various $V$$_\mathrm{G}$, where three of the data sets correspond to the data shown in Fig. 2(b). Paying attention to detailed changes in the current, we find two critical current values, illustrated as light green and blue dashed lines, which we denote as $I$$_\mathrm{1} \sim$ 65 pA and $I$$_\mathrm{2} \sim$ 80 pA. 
It should be noted that the value of $V$$_\mathrm{SD}$ at which the current increase from $I$$_\mathrm{1}$ to $I$$_\mathrm{2}$ occurs nearly corresponds to the crossing of the diamond borderline shown in Fig. 2(a) (see the pink, red, and blue arrows). If the FM lead was HMFs, the current would remain $I$$_\mathrm{1} \sim$ 65 pA in the whole region, by the spin blockade shown in Fig. 3(a). However, we actually use a Ni lead in the presence of the minority spin band at the Fermi level. In this context, we suggest that the current increases can be associated with ideal current jumps from the transition from one-electron tunneling to two-electron tunneling, and can be regarded as a consequence of the presence of minority spin transport. We can roughly estimate the contribution of the minority spin transport to the enhanced current to be ($I$$_\mathrm{2}$ - $I$$_\mathrm{1}$)/$I$$_\mathrm{1}$ $\sim$ 0.23, where $I$$_\mathrm{1}$ reflects the complete spin blockade in the two-electron tunneling regime. This value is nearly consistent with the values of 0.25--0.4, estimated from the spin polarization of Ni electrodes, $P$$_\mathrm{Ni}$ = 0.2--0.5.\cite{Soulen} 

Next, we focus on the detailed features for reverse bias. Shown in Fig. 4(b) is an enlarged figure of the $I$--$V$$_\mathrm{SD}$ characteristics for various $V$$_\mathrm{G}$ (near $V$$_\mathrm{G} =$ 1.05 V and 1.30 V). Taking into account Fig. 2(a), we can regard the absolute current ($\sim$250 pA), indicated by the black-arrow region in Fig. 4(b), as a consequence of one-electron tunneling. In the two-electron tunneling regime, an enhancement in the absolute current ($\sim$100 pA), indicated by the pink-arrow region, can also be observed. Since the ratio of the current increase can be estimated to be $\sim$ 0.4 ($\sim$100/$\sim$250) in the two-electron tunneling regime, we can interpret that this feature can also be attributed to the influence of minority spin transport.  
 
Finally, we comment on the sudden changes in $dI$/$d$$V$$_\mathrm{SD}$ in Fig. 2(a) at the conditions indicated by the black arrows. With increasing $V$$_\mathrm{SD}$ from the above two conditions of the current suppression, an influence due to the excited states ($\uparrow$, $\uparrow$), spin triplet and total spin $S =$ 1 on the current flowing can be considered. If the spin triplet is related to the electron transport, the spin blockade discussed above cannot be realized for the two-electron QD regime in this NM--QD--FM system. Further understanding of the effect of the excited states on the transport features is required. 

In summary, we have fabricated a semiconductor-QD spin-diode structure and examined its electron transport features. We detected anomalous suppression of the current in a two-electron QD regime. We discussed possible origins of the current suppression induced by spin blockade for the ground state of the two-electron QD. 

\begin{figure*}
\includegraphics[width=17cm]{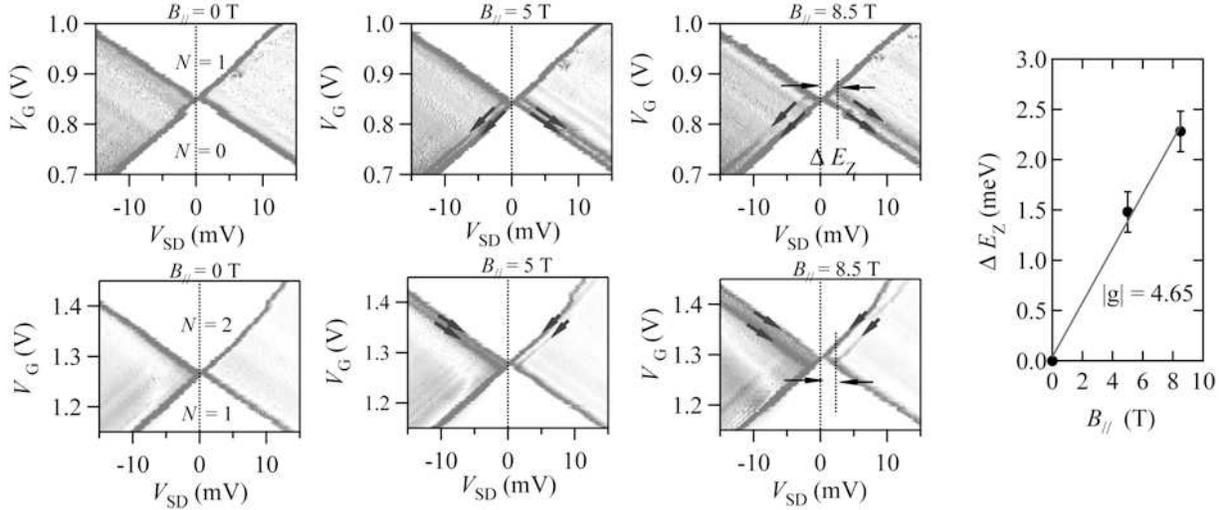}
\caption{[Appendix Figure] Coulomb stability diamond (CD), in the few-electron QD regime, measured at various applied magnetic fields.The energy splittings for both bias regimes can be clearly seen, as described in Ref.23.}
\end{figure*} 

\begin{figure}[t]
\includegraphics[width=8cm]{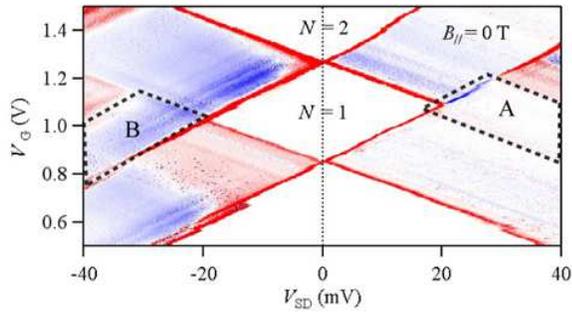}
\caption{[Appendix Figure] Differential conductance $dI$/$d$$V$$_\mathrm{SD}$ as a function of $V$$_\mathrm{SD}$ and $V$$_\mathrm{G}$ measured at $B =$ 0 T. }
\end{figure}

%\noindent{REFERENCES}


\begin{thebibliography}{11}
\bibitem{ref}
When the color scale is tuned, the energy splittings for reverse bias regimes can be clearly seen.
\end{thebibliography}
\end{document}